\documentstyle[11pt,fleqn]{article}

\newcommand{\R}{\mbox{$I\!\!R$}}             

\begin{document}


\hfill{\sl preprint - UTF 395 \\ hep-th/9701099 }
\par
\bigskip
\par
\rm


\par
\bigskip
\begin{center}
\bf
\LARGE
Geometric Entropy and Curvature Coupling in Conical Spaces:
 $\zeta$ Function Approach
\end{center}
\par
\bigskip
\par
\rm
\normalsize



\begin{center}\Large

Valter Moretti \footnote{e-mail:
\sl moretti@science.unitn.it}

\end{center}



\begin{center}
\large
\smallskip

 Dipartimento di Fisica, Universit\`a di Trento
\\and

\smallskip

 Istituto Nazionale di Fisica Nucleare,\\
Gruppo Collegato di Trento,\\ 38050 Povo (TN)
Italia

\end{center}\rm\normalsize



\par
\bigskip
\par
\hfill{\sl January 1997}
\par
\medskip
\par\rm



\begin{description}
\item{Abstract: }
\it \\
The local $\zeta-$function approach is implemented to regularize
the natural path integral definition of the geometric entropy in
 the large mass black hole
Euclidean manifold. The case of a massless field coupled with the 
(off-shell) singular curvature is considered. It is proved that the geometric
entropy is independent of  the curvature 
coupling parameter $\xi$ avoiding negative
values obtained in other approaches.
\par
\end{description}
\rm


\smallskip
\noindent{\sl PACS number(s):\hspace{0.3cm}
04.62.+v, 04.70.Dy}
\par
\bigskip
\rm



\section*{Introduction}

This short 
paper is devoted to develop a possible definition of the geometric 
entropy, employing the  path integral and the 
$\zeta$-function regularization in the case
of a scalar field propagating in a Euclidean manifold containing  conical
singularities, taking care to consider any possible choice of the 
 coupling  parameter $\xi$ with the
conical curvature. 

The geometric entropy in such  manifolds has been considered in many papers
due to its relation to the black hole entropy \cite{2,3,4,5,6,7,8,ffz,hotta}
and  the interesting remark that ultraviolet divergences exactly coincides 
with the divergences of the gravitational constant  in 
perturbation theory of quantum gravity \cite{jacobson,susskind,ffz} at the
Hawking-Unruh temperature.

Unfortunately,  some severe difficulties in developing these topics have been
pointed out as far as the non-minimal coupling is concerned \cite{susskind,
barbon,demers,hotta}. In that case, the geometric entropy seems to
  become
negative whenever $\xi >1/6$. 
However,  the geometric entropy, which is related to a completely 
(Ricci-) flat
manifold,
 seems to be dependent on $\xi$, the coupling  parameter  with the
Euclidean manifold (vanishing!) curvature.

A  recent  proposal  to improve these unwelcome
 results \cite{hotta}\footnote{We stress that  similar results to those 
found by
Hotta et. al in \cite{hotta} already appeared, in a generalized version,
within a  previous paper by 
Frolov, Fursaev and Zelnikov \cite{ffz}.} consists  in
redefining the geometric entropy by subtracting 
 a contribution 
arising from the ``off-shell'' singular curvature 
in
the formula proposed by Larsen and Wilczek \cite{larsen}.
This anomalous contribution 
 remains as a relic in the completely flat manifold.
On shell (namely at the Hawking-Unruh temperature $\beta^{-1}_H$),
this contribution can be represented in terms of a regularized
coincidence limit of the Euclidean Green function on the conical space
\cite{ffz,hotta}.

We remark that
 the calculations in \cite{ffz,hotta} have been performed through heat-kernel
regularization which involves, off shell, a non-Planckian behaviour of the
thermodynamical quantities expected by Susskind and Uglum in \cite{susskind}
at high temperatures.
 Rather, the local 
$\zeta$-function approach implemented
in the conical manifold \cite{zerbini} or in the optical related manifold
\cite{ielmo,moiel} avoids this drawback also considering photons and 
gravitons\footnote{The approach in the optical manifold completely
agrees with results based on point splitting procedure but  does not
 completely
agrees with the results arising
in the conical manifolds due to non-Planckian terms
relevant just at low temperatures.}.  

In this letter, we shall point out that  the local $\zeta$-function
  procedure to regularize the 
definition of geometric entropy (defined through  the usual
 path integral approach directly
 in
 the conical manifold) does overcome also the difficulties
related to the unexpected $\xi$ dependence in the geometric entropy. 
In fact, we shall see that the (on-shell) geometric entropy  resulting
from the {\em local} $\zeta$ function approach is  
 completely independent of the coupling with the singular curvature, and 
thus no problems arise with negative entropies, too.
 This holds in the case of a large mass Schwarzschild black hole
represented by the Euclidean Rindler manifold at least.

\section{Path integral approach}

Let us consider a four dimentional Euclidean manifold ${\cal M}_\beta =
{\cal C}_\beta \times
 \R^2$ endowed with the metric:
\begin{eqnarray}
ds^2 = -r^2d\theta^2 + dr^2 + dx_\perp^2 \label{conic},
\end{eqnarray}
where $x_\perp = (x^1, x^2 )$ are the flat transverse coordinates,
$\theta\in  [0,\beta]$ $(\beta \equiv 0)$,
 $r\in [0, +\infty)$, $x_\perp \in \R^2$, 
${\cal C}_\beta$ is a cone with angular deficit $0 <\beta \leq 2\pi$. 
${\cal M}_\beta $ takes on a singular curvature
 \cite{conic}\footnote{The delta function is defined 
in such a manner that 
\begin{eqnarray}
\int_{{\cal C}_\beta} \sqrt{g} \delta(r) d^2x =1 \nonumber. 
\end{eqnarray}}:
\begin{eqnarray} 
R(x) = 4\pi (1- \beta/2\pi)
\delta(r)
\end{eqnarray}
We remark that this  singular 
 curvature vanishes at $\beta=\beta_H=2\pi$.
The above-considered manifold is the Euclidean section of the spacetime near
the horizon of a Schwarzschild black hole (Euclidean Rindler space)
or, equivalently, it is the Euclidean 
manifold corresponding to a very large mass
black hole.  
The particular value $\beta_H =2\pi$ defines the inverse Hawking-Unruh
temperature related to the thermodynamical equilibrium quantum state
of the fields propagating around the black hole.

Following \cite{larsen} and \cite{hotta}, we compute the geometric 
entropy of  a massless minimally coupled field employing a path integral as:
\begin{eqnarray}
S = \left(1-
 \beta \frac{d\:\:}{d \beta}
\right)|_{\beta=2\pi}\int {\cal D}\phi \: e^{-A_{\beta}[\phi]}, 
\label{geometric}
\end{eqnarray}
where $A_\beta$ is the Euclidean action of the field in our time-periodic
manifold. The previous relation is formally equivalent to the usual 
definition of the thermodynamic entropy within the canonical 
ensemble approach. Anyhow, there is an important
 difference. One can check the  formula above-written without to  employ 
thermodynamical laws (and thus avoiding possible troubles with ill-defined
thermal states off-shell) by assuming the formal ``geometric'' 
 and non-termodinamical definition:
 \begin{eqnarray}
S = - \mbox{tr} (\rho \ln \rho) \nonumber 
\end{eqnarray}
and only then proving  (\ref{geometric}) as pointed out in \cite{hotta}.
Above, $\rho$ is the statistical operator representing the global 
Hartle-Hawking-Minkowski vacuum confined in a Schwarzschild-Rindler wedge.
The functional integral is then interpreted as the determinant of the 
differential  operator: 
\begin{eqnarray}
L = -\nabla_a\nabla^{a}+ m^2
\end{eqnarray}
appearing in the Euclidean action:
\begin{eqnarray}
A_\beta[\phi] = -\frac{1}{2}\int d^4x \sqrt{g}
\phi [ \nabla_a\nabla^{a}- m^2]\phi
\label{minimal}.
\end{eqnarray}
As usually,  we get the identity:
\begin{eqnarray}
\int {\cal D}\phi e^{-A_{\beta}[\phi]} = \left[ \det (-\nabla_a\nabla^{a}
+m^2) \right]^{-1/2}  \nonumber.
\end{eqnarray}
Finally, introducing the regularization 
scale $\mu$ (which  disappears in the final 
formulae in our case due to $\zeta(0|L)=0$)
 we can compute the previous determinant in the framework of the
 $\zeta$ function 
regularization \cite{hawking} as:
\begin{eqnarray}
\ln \det [(-\nabla_a\nabla^{a} +m^2)/\mu^2] = -\frac{d}{ds}|_{s=0}
\zeta(s|L) -2\zeta(0|L)\ln\mu \nonumber.
\end{eqnarray}  
The $\zeta$ function can be obtained by integrating
 the {\em local} $\zeta$ function:
\begin{eqnarray}
\zeta(s|L) = \int d^4x\sqrt{g} \zeta(s,x|L), \label{int}
\end{eqnarray}
where, through the spectral representation, $\phi_n(x)$ being the 
normalized eigenvector
of $L$ with eigenvalue $\lambda_n$:
\begin{eqnarray}
\zeta(s,x|L) = \sum_n \lambda_n^{-s} \phi_n(x)\phi_n(x) \label{locale}.
\end{eqnarray}
This identity has to be understood in the sense of the analytic continuation
  of the right hand side to values of $s$
 by which the summation does not converge.
Usually, for Euclidean 
compact 4-manifolds, the summation above converges whenever
 $ \mbox{Re}$ $s > 2 $ defining an analytic  function which can be extended 
on the whole complex $s$ plane except  for simple poles on the  
real axis; we refer to \cite{zerbinil} for a complete report in the 
general case.\\
Another very important feature of the local $\zeta$ function approach
 is that the value $\zeta(1,x|L)$, whenever it exists,
gives rise to a natural regularized definition of the Euclidean Green 
function, 
corresponding to the eigenfunctions $\phi_n$  of the operator $L$,
in the case of coincidence 
of  arguments. This trivially follows from (\ref{locale}). 
  
All the calculations so far considered have been explicitly performed 
in the case $m=0$  for the metric in (\ref{conic}) (including
several 
generalizations) by Zerbini et al. in \cite{zerbini}, employing the local
$\zeta$ function approach.
The local $\zeta$ function has been proved to read:
\begin{eqnarray}
\zeta(s,x |L) = \frac{r^{2s-4}}{4\pi\beta \Gamma(s)} 
I_\beta(s-1) \label{zeta}
\end{eqnarray}
where the fuction $I_\beta(s)$ is analytic on the whole complex $s$ plane 
but in $s=1$, where it has a simple pole with residue $\frac{1}{2}
[(\beta/2\pi)-1]$; other important values are:
\begin{eqnarray}
I_\beta(0) = \frac{1}{6}\frac{\beta}{2\pi}
\left[\left(\frac{2\pi}{\beta} \right)^2 -1 \right] \label{i0}
\end{eqnarray}
and
\begin{eqnarray}
I_\beta(-1) = \frac{1}{90}\frac{\beta}{2\pi}
\left[\left(\frac{2\pi}{\beta} \right)^2 -1\right]
\left[\left(\frac{2\pi}{\beta} \right)^2 +11\right] \nonumber .
\end{eqnarray} 
The corresponding  geometric entropy reads:
\begin{eqnarray}
S_{\beta=2\pi} =  \frac{A_{\perp}\beta}{720 \pi^2 \epsilon^2}
\left[\left(\frac{2\pi}{\beta}\right)^4 + 5
\left(\frac{2\pi}{\beta} \right)^2 
\right] \mid_{\beta=2\pi} = \frac{A_{\perp}}{60 \pi \epsilon^2}
\label{entropy}.
\end{eqnarray}
Notice the Planckian high temperature behaviour at $\beta \neq \beta_H =2\pi$
expected from statistical mechanics \cite{susskind}\footnote{We stress 
that also the high temperature coefficient in the above formula 
coincides exactly  with
 the corresponding coefficient expected in \cite{susskind} from
statistical mechanics.}
and the cutoffs $A_\perp$ and $\epsilon$,
 the area of the event horizon and  
the minimal distance from the horizon respectively,
 necessary in computing the integral in (\ref{int}).
As far as the arguments coincidence limit of the Euclidean Green function 
in the considered ${\cal C}_\beta\times\R^2$ space is concerned,
 few calculations, employing the explicit form of the well-known 
Green function
(e.g. see \cite{mova} and ref.s therein),  prove that 
the regularized definition of 
$<\phi(x)\phi(x)>_\beta $: 
\begin{eqnarray}
<\phi(x)\phi(x)>^{\scriptsize \mbox{regularized}}_\beta 
= \zeta(1,x| L)
\end{eqnarray}
  is equivalent to the natural
divergence subtraction procedure:
\begin{eqnarray}
<\phi(x)\phi(x)>^{\scriptsize \mbox{regularized}}_\beta 
= <\phi(x)\phi(x)>_\beta - <\phi(x)\phi(x)>_{\beta=2\pi}\nonumber. 
\end{eqnarray}  
Following our regularization, we have:
\begin{eqnarray}
<\phi(x)^2>_\beta = \frac{1}{48\pi^2 r^2} 
\left[\left(\frac{2\pi}{\beta} \right)^2 -1\right]
\end{eqnarray}
In particular, not depending on the  value of
$x_\perp$, introducing the usual ultraviolet cutoff $\epsilon$ near the horizon,
the natural definition of the averaged square field on the horizon
reads: 
\begin{eqnarray}
<\phi(0)^2>_\beta = \frac{1}{48\pi^2 \epsilon^2}
\left[\left(\frac{2\pi}{\beta} \right)^2 -1\right] \label{phizero}.
\end{eqnarray}
Notice the simple zero at $\beta=2\pi$.

\section{General coupling}

The interesting problem is now to investigate whether or not the considered
approach makes sense also in the case of a general coupling with the 
singular curvature in the Euclidean Lagrangian: $ \frac{1}{2}
 \xi R \phi^2$.\\
We shall investigate the case $m=0$ only. In this case we are able to
 perform
explicit calculations through the  $\zeta$ function approach.
Following \cite{larsen}, let us  define once again:
\begin{eqnarray}
S(\xi) = \left(1-
 \beta \frac{d\:\:}{d \beta}
\right)|_{\beta=2\pi}\int {\cal D}\phi \: e^{-A_{\xi\beta}
[\phi]} = \left(1-
 \beta \frac{d\:\:}{d \beta}
\right)|_{\beta=2\pi} Z_\beta(\xi),  \label{def}
\end{eqnarray}
where:
\begin{eqnarray}
A_{\xi\beta}[\phi] = -\frac{1}{2}\int d^4x \sqrt{g}
\phi [ \nabla_a\nabla^{a}- \xi R \phi^2]\phi 
\label{general}.
\end{eqnarray}
The heat kernel approach, implemented in this case  to regularize
the determinant following from the path integral written above,
produces a $\xi$-dependent
 entropy which is negative when $\xi > 1/6$ \cite{larsen,hotta}.
This is the fundamental reason of Hotta et al. 
\cite{hotta} to reject the previous definition 
 and propose an improved, not natural in our opinion,  definition.
Let us prove that, following the local $\zeta$ function approach
and implementing an analysis similar to that in \cite{hotta}, the 
definition  (\ref{general}), produces a geometric entropy which does not
depend on $\xi$; we stress that this holds only
``on shell'', namely when $\beta = \beta_H$, as one could expect.

Trivial manipulations of the formulae above leads us to:
\begin{eqnarray}
Z_\beta(\xi) = Z_\beta(0) <e^{-2\pi\xi (1-\beta/2\pi)   
\int dx_\perp \phi(0,x_\perp)^2 }>_\beta
\end{eqnarray}
where the average has been taken with respect to the minimal coupling. 
Notice also that:
\begin{eqnarray}
<e^{-2\pi\xi (1-\beta/2\pi)\int dx_\perp \phi(0,x_\perp)^2 }|_{\beta=\beta_H}  
>_\beta = <1>_\beta = 1
\end{eqnarray}
not depending on $\beta$.\\
Taking account of the identity above  as well as the $\zeta$
 regularized definition in (\ref{phizero}), by
 employing the definition (\ref{def}), we have:
\begin{eqnarray}
S(\xi) &=& S(\xi=0) + 2\pi Z_{2\pi}  \frac{d\:\:}{d \beta}
|_{\beta=2\pi} <1>_\beta + 2\pi <
\frac{d\:\:}{d \beta}|_{\beta=2\pi}   e^{-2\pi\xi (1-\beta/2\pi)
\int dx_\perp \phi(0,x_\perp)^2 }>_{2\pi} \nonumber \\
&=& S(\xi=0) + 2\pi \xi Z_{2\pi} A_{\perp} <\phi(0)^2>_{2\pi} =  S(\xi =0) 
\end{eqnarray}
We see that, regularizing the theory by the local $\zeta$ function
approach, the geometric entropy does not depend on $\xi$ 
and coincides to that evaluated in the minimal
coupling.
We stress that this result holds just when
$\beta=\beta_H (=2\pi)$ as one could expect naively.

\par \section*{Acknowledgments}
I would like to thank   Luciano Vanzo 
and Sergio Zerbini for some helpful  discussions.
I am also grateful to D.V. Fursaev who pointed out 
the ref. \cite{ffz} to me.

\end{document}